\documentclass[letterpaper,twocolumn,10pt]{article}
\usepackage{usenix}

\usepackage{amsmath,amssymb}
\usepackage{graphicx}
\usepackage{xcolor}
\usepackage{booktabs}
\usepackage{url}
\usepackage{subcaption}
\usepackage{tikz}
\usepackage{xspace}
\usepackage{hyperref}
\hypersetup{hidelinks}

\providecommand{\Description}[1]{}

\providecommand{\fcirc}[1]{
  \tikz[baseline=(char.base)]{
    \node[shape=circle, fill=black, inner sep=1.2pt] (char) {\color{white}\scriptsize #1};
  }
}
\providecommand{\sys}{Tessera\xspace}

\date{}
\title{\sys: Demand-Driven KV Cache Management for Retrieval-Augmented LLM Serving}
\author{{\rm Fei Fang\thanks{These authors contributed equally.} \quad
 Chung-Hsiang Lo\footnotemark[1] \quad
 Yi Liu \quad Yifan Hua \quad Chen Qian}\\[0.35em]
University of California, Santa Cruz
}

\begin{document}
\maketitle

\begin{abstract}
RAG and retrieval-based agent memory both inject retrieved content into LLM prompts, as document chunks and recalled memory records, respectively. The same content can recur across requests at different prompt positions or after different preceding contexts, preventing reuse through conventional prefix caching. Our characterization finds that records recurring outside the matching prefix account for over 70\% of injected memory tokens in agent-memory workloads. Composable KV-reuse methods enable reuse in such cases, but online serving introduces a management problem: a recurring unit's KV states may not yet exist, may have been evicted, or may reside on another node. We present \sys, a disaggregated serving system that makes retrieval the control plane for KV reuse. By exposing the context units needed before model execution, retrieval allows \sys to combine current demand with retrieval history, KV residency, and generation load to coordinate cache management and request routing. Generation nodes concurrently prepare locally cached, remotely cached, and missing states, while retaining newly computed states off the request's critical path. Across RAG and agent-memory workloads, \sys lowers mean TTFT by up to 3.6$\times$ over SGLang and LMCache with EPIC at matched request rates, and sustains low TTFT at rates where the baselines saturate, while matching the answer quality of the underlying composition policy.
\end{abstract}

\vspace{-4ex}
\section{Introduction}
\label{intro}
\begin{figure}[t]
    \centering
    \includegraphics[width=1\columnwidth]
    {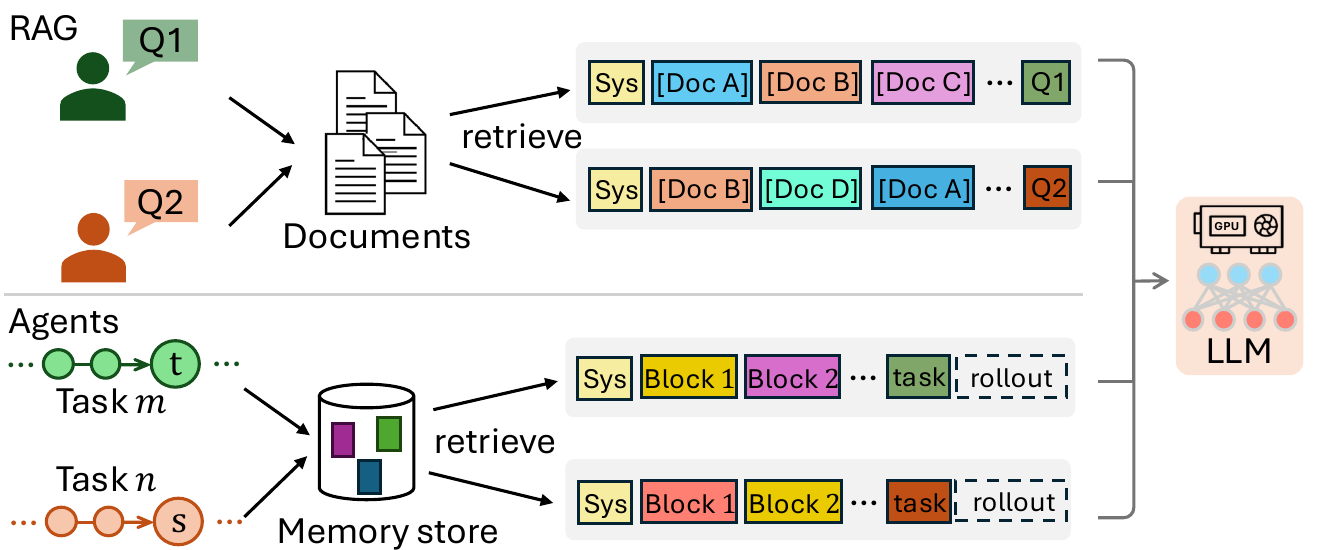}
    \vspace{-4ex}
    \caption{Retrieved context units in RAG and retrieval-based agent memory.}
    \vspace{-5.5ex}
    \label{fig:intro:workloads}
\end{figure}

\vspace{-3ex}
Large language model (LLM) applications 
construct prompts by retrieving text from external state. Retrieval-augmented generation (RAG) retrieves document chunks from a knowledge base to ground generation in external evidence~\cite{gao2023retrieval,guu2020retrieval,lewis2020retrieval}. Similarly, agents with retrieval-based memory recall records of previous interactions, observations, or experiences when constructing prompts for subsequent steps~\cite{packer2023memgpt,zhang2025memorysurvey}. Although these applications retrieve different types of content, they share a common input structure: each request combines independently selected text segments with task-specific instructions and a query for a single model invocation. Figure~\ref{fig:intro:workloads} illustrates this pattern in both settings, with retrieved document chunks or memory records arranged as an ordered sequence within the prompt. We use \emph{context unit} to denote each independently retrieved segment, including a document chunk in RAG or a memory record in an agent system.

Before generating a response, the serving engine must prefill the request prompt and construct key and value (KV) states for both the request-specific text and its retrieved context units. For requests with long retrieved contexts, this computation can dominate time to first token (TTFT) and consume substantial GPU resources~\cite{jin2025ragcache}. Much of this work is repeated. Popular document chunks recur across RAG queries, while memory records may be recalled by multiple steps or sessions. Conventional prefix caching cannot capture reuse when the same context unit appears at a different prompt position or after a different preceding context. This pattern occurs in both RAG and retrieval-based agent memory. In our two agent workloads, repeated memory records outside the matching prefix account for 73.9\% and 76.3\% of injected memory tokens (Section~\ref{sec:agent_workload}).

In RAG, the dependence of a chunk's KV states on its prompt position and preceding context has motivated composable chunk-level KV reuse, which prepares chunks independently and assembles their cached states at inference time~\cite{gim2024prompt,lu2025turborag,yao2025cacheblend,hu2025epic,yang2026kvlink}. Existing methods address positional placement and cross-chunk context through mechanisms such as position-preserving layouts, selective recomputation, modified attention, or learned link tokens. These techniques make it possible to reuse a recurring unit beyond the matching prefix. They do not, however, ensure that its states are available when it is retrieved again: under dynamic retrieval demand and finite cache capacity, a unit's states may never have been computed, may have been
evicted, or may reside on a node other than the one serving the request. 

RAGCache~\cite{jin2025ragcache} stores prefix-dependent document states in a knowledge tree, while Cache-Craft~\cite{agarwal2025cache} maintains context-specific chunk variants with selective recomputation. Their cache-management designs focus on node-local memory hierarchies. Distributed KV systems support cross-node cache access and cache-aware scheduling~\cite{qin2024mooncake,liu2025lmcache, ICLR2025_5bc342f4}, but their decisions are not driven by the retrieval demand that determines which units will recur. Turning recurrence into reuse in online serving raises three challenges.  First, cache management and request routing are coupled. Where a unit's states are retained determines which node can reuse them, while routing determines where that demand arrives: sending a request to the node holding most of its states can overload that node, whereas dispatching by load alone forfeits local reuse. Second, a request may mix cached and missing states, some of which must be transferred while others must be computed from text. Preparing them serially accumulates both delays on the request's critical path, while blocking the scheduler until all are ready delays unrelated requests. Third, retaining newly computed states benefits only future requests but competes with the current one for memory and transfer
resources. Retention can be declined when capacity is short or reuse value is low, yet the current request must still be served. Our key insight is that retrieval exposes the context units a request needs before model execution, making the retrieval tier a natural control plane for KV reuse. Retrieval alone does not know where those units' states reside or how valuable they are; combined with the demand history, KV residency, and generation load that the serving system maintains, however, this per-request demand lets the control plane decide cache retention and request routing before dispatch, addressing the first and third challenges. The same unit identities tell the selected node which states to fetch and which to compute, so cached-state access and miss prefill can proceed together,
addressing the second. Because retention is decided separately from preparation, a request proceeds once its states are ready, even when retention is declined or still in progress.

We present \sys, a serving system that realizes this design for RAG and retrieval-based agent memory. \sys disaggregates retrieval, shared KV storage, and generation into independently provisioned pools. The Retrieval Pool serves as the control plane, tracking context-unit demand and residency and jointly deciding cache retention and cache-aware routing. The shared KV Pool extends reuse across generation nodes. Generation nodes prepare cached
and missing states concurrently, compose them before completing prefill, and retain authorized states in the background. Our prototype composes states with EPIC's LegoLink policy~\cite{hu2025epic}. Other policies that operate on independently prepared unit states can use the same interfaces.

In summary, this paper makes the following contributions:
\begin{list}{\textbullet}{%
    \setlength{\leftmargin}{1.2em}%
    \setlength{\labelwidth}{0.8em}%
    \setlength{\labelsep}{0.4em}%
    \setlength{\topsep}{1pt}%
    \setlength{\itemsep}{0pt}%
    \setlength{\parsep}{0pt}%
    \setlength{\partopsep}{0pt}%
}
    \item We characterize retrieval-based agent memory and find that 73.9--76.3\% of injected memory tokens recur beyond the matching prompt prefix, exposing substantial reuse unavailable to conventional prefix caching.
    \item We present \sys, a disaggregated serving system that couples retrieval demand with KV residency and generation load to coordinate cache management and request routing, while supporting mixed cached/missing execution and nonblocking state retention.
    \item We implement a \sys prototype by extending SGLang~\cite{zheng2024sglang}. At matched request rates, \sys lowers mean TTFT by up to 3.6$\times$ over SGLang and LMCache~\cite{liu2025lmcache} with EPIC, and sustains low TTFT where the baselines saturate, while matching the answer quality of the underlying composition policy.
    \item We release the \sys prototype and, to our knowledge, the first public request-level traces of agents with retrieval-based memory together with the memory stores they recall from.
    \end{list}

\vspace{-4ex}
\section{Background and Motivation}
\label{bg}
\vspace{-2.5ex}

We review prefix caching (Section~\ref{bg:infer}) and composable KV reuse for retrieved context in RAG (Section~\ref{bg:rag}) and retrieval-based agent memory (Section~\ref{bg:agent}). We then examine what changes when composable KV reuse moves to online serving, and the challenges this raises (Section~\ref{bg:problem}).

\vspace{-3ex}
\subsection{LLM Inference and Prefix KV Caching}
\label{bg:infer}
\vspace{-2ex}
Most modern LLM serving engines run autoregressive Transformer models~\cite{vaswani2017attention, chowdhery2023palm}. Given an input prompt, such a model first performs a \emph{prefill} pass to process all input tokens and compute their corresponding KV states, which are stored in the request's \textit{KV cache}. Each decode step then attends to these cached states rather than recomputing them for earlier tokens. For long prompts, constructing these states can make prefill a major contributor to TTFT. Serving engines such as vLLM and SGLang reduce repeated prefill through prefix KV caching~\cite{kwon2023efficient,zheng2024sglang}, but only when requests share an identical prompt prefix. Because a token's KV states depend on its position and on all preceding tokens, reuse stops at the first divergence: identical text that follows a different prefix cannot use the cached states directly.

\vspace{-3.5ex}
\subsection{RAG Retrieval and Composable KV Reuse}
\label{bg:rag}
\vspace{-2ex}
\noindent\textbf{Retrieval.} RAG enhances LLMs by incorporating retrieved external knowledge into generation. Retrieval may use vector search~\cite{laskar2020query,yao2017recent}, graph-based search~\cite{edge2024local,sarthi2024raptor,guo2024lightrag,gutierrez2025rag}, or combinations of the two. We consider pipelines whose output reaches the LLM as an ordered sequence of textual context units. Because retrieval is query-dependent, the selected units and their order vary across requests, so the same unit can appear at different positions in different prompts.
\begin{figure}[t]
    \centering
    \includegraphics[width=1\linewidth]{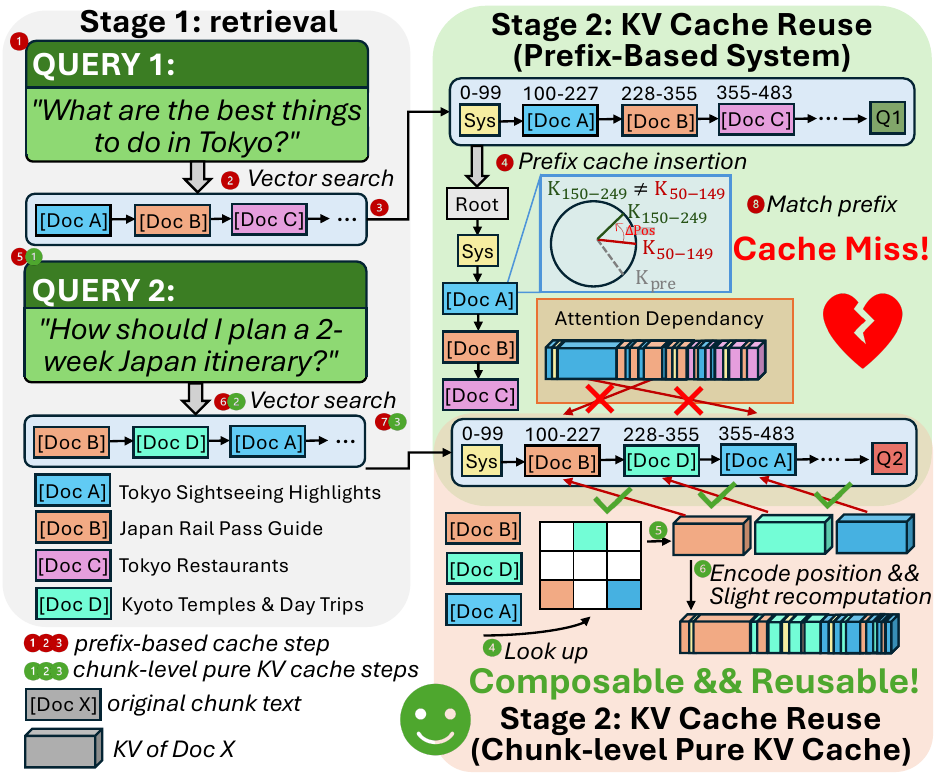}
    \vspace{-5ex}
    \caption{Dynamic RAG retrieval vs. prefix caching.}
    \vspace{-5.5ex}
    \label{fig:intro:mov1}
\end{figure}

\noindent\textbf{Limitations of prefix KV reuse.}
Figure~\ref{fig:intro:mov1} gives an example. Suppose Q1 retrieves \texttt{[Doc A, Doc B, Doc C]} and Q2 retrieves \texttt{[Doc B, Doc D, Doc A]}. \texttt{Doc B} occupies positions 150--249 after \texttt{Doc A} in Q1, but positions 50--149 with no preceding document in Q2. Its cached keys therefore carry rotary position embedding (RoPE) rotations for the wrong positions, and
its keys and values encode a dependency on \texttt{Doc A} that no longer holds. Conversely, the states cached for \texttt{Doc A} in Q1 lack the dependencies on \texttt{Doc B} and \texttt{Doc D} that precede it in Q2.
The text is unchanged, yet a prefix cache must prefill both documents again.

\noindent\textbf{Composable Context Unit KV Reuse.}
This mismatch motivates treating context units as independently prepared reuse units. Recent methods extend reuse beyond matching prefixes by preparing context units independently and composing their cached states at inference time~\cite{gim2024prompt, lu2025turborag, yao2025cacheblend, hu2025epic, yang2026kvlink}. Such composition must address both positional placement and cross-unit context. We target RoPE-based models~\cite{qwen3technicalreport,gemmateam2024gemmaopenmodelsbased}, which rotate queries and keys but not values; caching keys before RoPE lets them be rotated for a unit's new offset.
This removes the keys' dependence on absolute offsets but not on preceding context.
When a context unit is prefilled independently, its deeper-layer KV states do not incorporate causal influences from preceding units in the assembled prompt. Existing methods provide different mechanisms for positioning and cross-unit context, including position-preserving layouts, position restoration, selective recomputation, modified attention, and learned link tokens~\cite{gim2024prompt,lu2025turborag,yao2025cacheblend,hu2025epic,yang2026kvlink}. These approaches offer different quality--computation tradeoffs and do not provide identical composition semantics. The composition rule is an algorithmic choice separate from the online cache management problem studied in this paper. Accordingly, \sys separates the algorithm-specific composition stage from its common path for context-unit identification, cache management, state access, and miss execution. Our prototype instantiates this stage with EPIC's LegoLink policy~\cite{hu2025epic}.

\subsection{Retrieval-Based Agent Memory}
\label{bg:agent}
\vspace{-2ex}
\begin{figure}[t]
    \centering
    \includegraphics[width=1\linewidth]{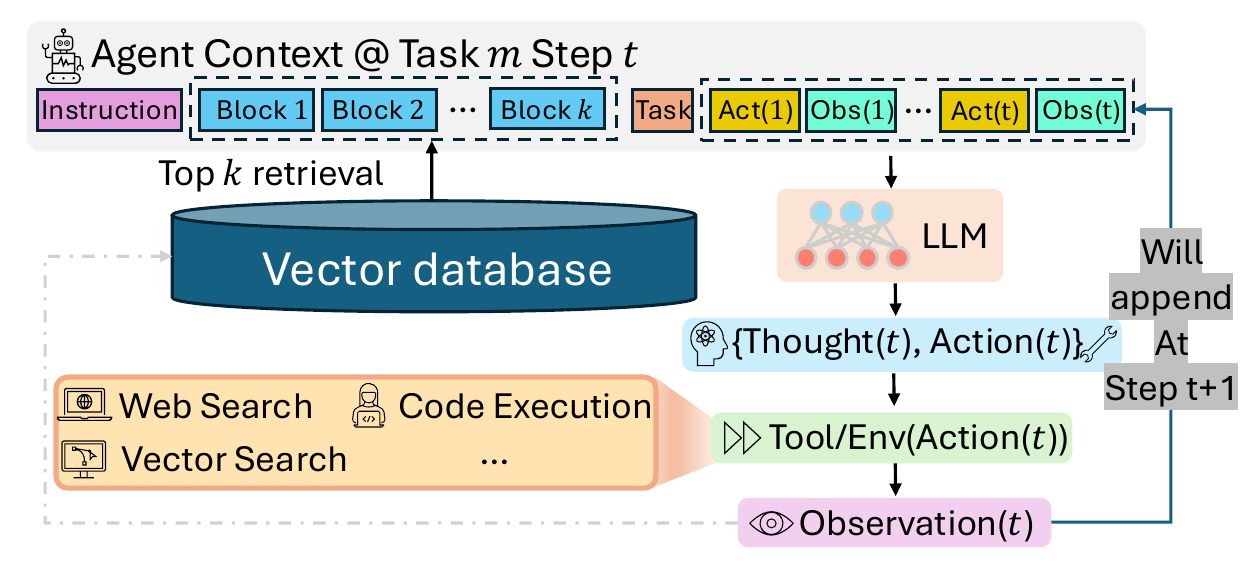}
    \vspace{-5ex}
    \caption{A typical retrieval-based memory workflow in an LLM agent.}
    \label{fig:bg:agent-workflow}
\end{figure}
\noindent\textbf{Memory retrieval across agent steps.} 
Retrieval-based agent memory retains information from prior interactions in external storage for later recall. Systems such as Mem0~\cite{doi:10.3233/FAIA251160} and OpenViking~\cite{openviking} extract, maintain, and retrieve memory records, supporting memory writing, management, and reading~\cite{zhang2025memorysurvey}. As illustrated in Figure~\ref{fig:bg:agent-workflow}, an agent execution forms a multi-step trajectory. At a memory-augmented step, the model input combines task instructions and recent observations with memory records retrieved according to the current task state.
The retrieved memory may originate from external knowledge, such as files, or information accumulated from prior interactions. The LLM may then invoke external tools or environments, producing new observations that extend the trajectory and influence subsequent steps. 

\begin{figure}[t]
    \centering
    \includegraphics[width=1\linewidth]{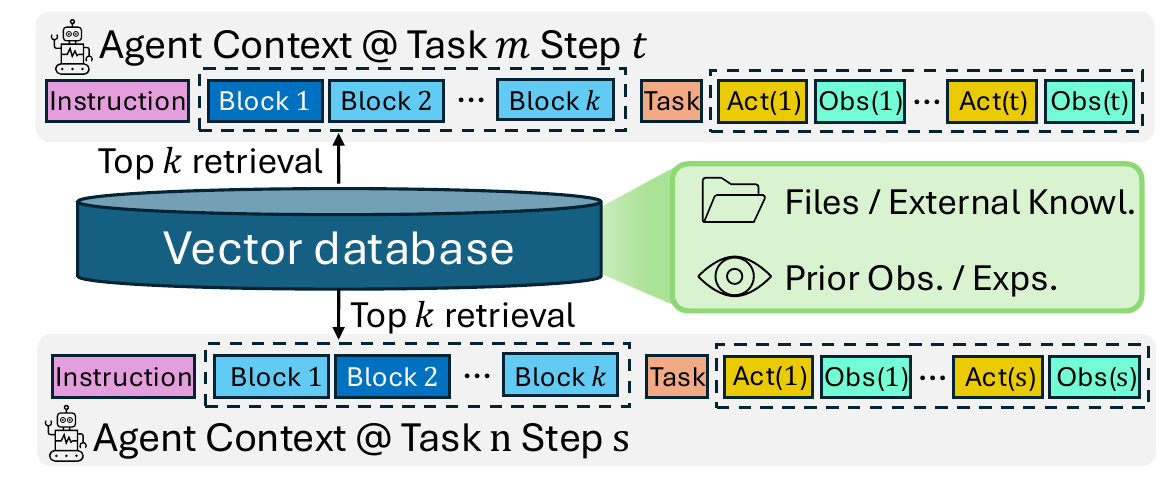}
    \vspace{-5ex}
    \caption{Non-prefix recurrence across agent trajectories. }
     \vspace{-3ex}
    \label{fig:intro:agent-recurrence}
\end{figure}

\noindent\textbf{Common interface, distinct workloads.}
Like retrieved documents in RAG, recalled memory records become part of the LLM prompt and require prefill. The workloads nevertheless differ in record lifecycles, sharing scopes, interaction histories, and recurrence patterns. We therefore focus on reusing unchanged records within their permitted sharing scope, leaving memory extraction, maintenance, and retrieval selection to the memory system.

\noindent\textbf{Recurrence within and across agent trajectories.}
The same memory record may be retrieved at nonconsecutive steps within an agent trajectory or across trajectories permitted access to it. In Figure~\ref{fig:intro:agent-recurrence}, the highlighted memory block appears as the first retrieved block at step $t$ of task $m$ and the second at step $s$ of task $n$. The additional preceding block changes its prompt position and preceding context despite its unchanged content. Prefix caching cannot exploit this recurrence once the preceding token sequences diverge. Section~\ref{sec:agent_workload} characterizes this non-prefix recurrence in agent workloads.

\begin{figure*}[t]
    \centering
    \begin{minipage}[t]{0.57\textwidth}
        \centering
        \includegraphics[width=\linewidth]
        {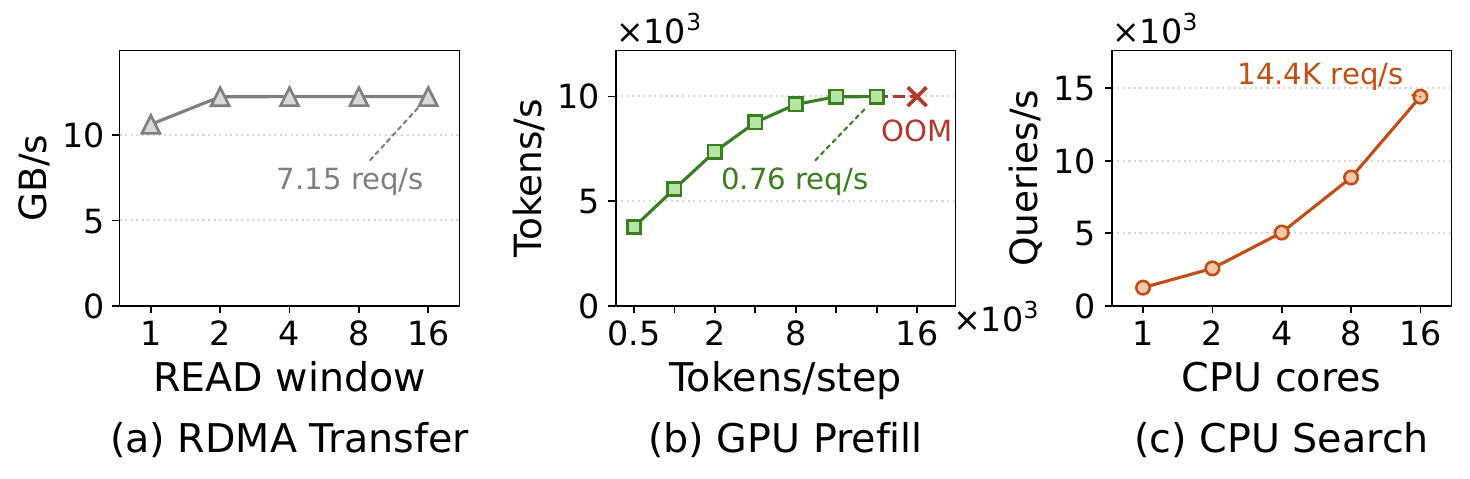}
        \vspace{-4.5ex}
        \caption{Capacity of vector search, KV fetch, and KV prefill.}
        \label{fig:component-capacity}
    \end{minipage}
    \hfill
    \begin{minipage}[t]{0.41\textwidth}
        \centering
        \includegraphics[width=\linewidth]{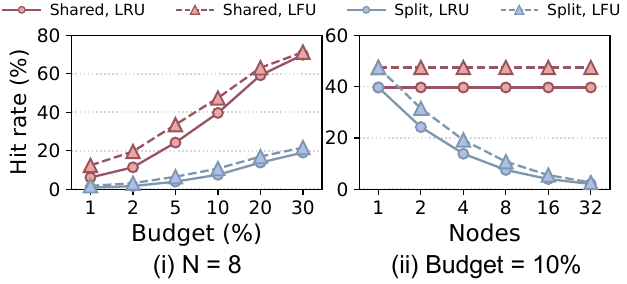}
        \vspace{-4.5ex}
        \caption{Chunk hit rate of a shared cache vs.\ $N$ split caches.}
        \vspace{-4ex}
        \label{fig:background-pooling}
    \end{minipage}
    \vspace{-2ex}
\end{figure*}

\subsection{From Composable KV Reuse to Online Serving}
\label{bg:problem}
\noindent\textbf{Problem: Online Composable KV Reuse Requires Demand-Driven State Management.}
The preceding sections illustrate how context units in RAG and retrieval-based agent memory can recur outside a matching prompt prefix. Section~\ref{sec:agent_workload} measures how often this occurs in agent workloads. We target online deployments in which requests are served across multiple generation nodes under bounded KV-cache capacity. Composable KV-reuse methods address how independently prepared states can be positioned and combined while recovering cross-unit context~\cite{yao2025cacheblend,hu2025epic}. These mechanisms, however, do not by themselves determine how reusable states should be managed in an online serving system. Dynamic retrieval continually changes the active set of context units, while finite cache capacity can prevent retaining KV states for an entire document corpus or memory store. Consequently, the recurrence of a context unit does not guarantee that its KV states are available when the unit is retrieved again. The central systems problem is therefore to convert context-unit recurrence into actual reuse: deciding which states to materialize and retain, making them accessible across generation nodes, and falling back efficiently to prefill when they are absent. Serving a request in this setting involves retrieval, KV fetch, and KV prefill, which rely on different resources and operate at substantially different rates. Figure~\ref{fig:component-capacity} reports their native throughput on a 200-request 2WikiMQA~\cite{ho-etal-2020-constructing} workload, together with their request-equivalent rates. These rates span more than four orders of magnitude, from 0.76 req/s for KV prefill to 14.4K req/s for retrieval. This disparity motivates provisioning the three stages independently.

\noindent\textbf{Challenge 1: Cache Management and Request Routing Are Coupled.}
In a distributed deployment, the value of retaining a context unit depends on both future retrieval demand and where subsequent requests execute. Pooling capacity improves reuse: Figure~\ref{fig:background-pooling} shows that, on a 2WikiMQA workload, splitting the same aggregate capacity across eight per-node caches reduces the hit rate by roughly $5\times$, under both LRU and LFU. Yet node-local caches avoid network transfers, creating a tradeoff between hit rate and access cost. Cache management and routing therefore cannot be optimized independently: admission and eviction determine where reusable states reside, while routing determines whether requests can exploit that locality. Effective reuse requires coordinating these decisions across local and shared caches while accounting for retrieval demand and generation load.

\noindent\textbf{Challenge 2: Cache Hits and Misses Must Execute Concurrently.}
After routing, a request may require both cached context-unit states that must be accessed from their storage locations and missing states that must be computed from text. These preparation paths consume different resources and complete at different times, while subsequent composition and query processing depend on their results. Serializing them places both transfer and miss-prefill latency on the critical path, whereas blocking the scheduler until all states are ready delays unrelated requests and reduces batching opportunities. The serving engine must therefore coordinate preparation work that becomes ready at different times while continuing to serve other requests.

\noindent\textbf{Challenge 3: Retention Competes with Request Execution.}
Retaining a newly computed context unit enables future reuse, but competes with the current request for host memory, registered buffers, and the same NIC path that remote hits use to fetch cached states. The cost is substantial: for $L$ layers, $H_{kv}$ KV heads, head dimension $d_h$, tensor-parallel degree $\mathrm{TP}$, and element width $w$, each token produces $2L(H_{kv}/\mathrm{TP})d_hw$ bytes of KV state per rank. For example, Qwen3-32B has $L{=}64$, $H_{kv}{=}8$, and $d_h{=}128$, so with 16-bit KV states each token of retained context occupies 64\,KiB per rank. NIC provisioning does not follow GPU count. Our servers pair four GPUs with a single NIC, so at TP$=$4 all four ranks share one link and every single retained token moves 256\,KiB across it. Synchronous retention therefore adds significant data movement to the request's critical path, while asynchronous retention introduces publication and lifetime races if incomplete or in-flight states can be read or reclaimed. Retention must therefore proceed off the critical path while publishing states only after completion and protecting storage used by in-flight transfers.

\noindent\textbf{Opportunity: Retrieval Exposes Context-Unit Demand Before Model Execution.}
When retrieval completes, the identities and order of the context units a request needs are already known. Retrieval does not inherently know their KV residency or future value, but the serving system can associate this per-request demand with retrieval history, cache residency, and generation load. Retrieval therefore provides a control point for deciding admission, eviction, and routing before model execution, and for telling the selected node how to obtain each required state.
\section{Agent Workload Characterization}
\label{sec:agent_workload}
\vspace{-2.5ex}
Previous work has shown that popular document chunks recur across RAG queries~\cite{jin2025ragcache}, but whether agent memory offers the same reuse opportunity is unclear. The key question is not whether memory records recur, but whether they recur beyond the matching prompt prefix: repeats within the prefix are already captured by prefix caching, while those beyond it represent unexploited reuse. We measure this distinction on two agent workloads and find that non-prefix recurrence is prevalent in both.
\vspace{-1ex}
\subsection{Workloads and Methodology}
\vspace{-1ex}
\label{sec:agent_workload:method}
\noindent\textbf{Workloads and trace collection.}
We collect prompt traces from OpenClaw~\cite{openclaw}, an open-source agent
framework, using OpenViking~\cite{openviking} as its retrieval-based memory
store. A \emph{memory-augmented request} is the first LLM request following a
memory injection, retained with its complete rendered prompt.
LoCoMo~\cite{maharana2024locomo} is a long-term conversational-memory benchmark of questions
about facts stated in long conversations between recurring speakers. We import
four conversations into a shared OpenViking service as four isolated
user--agent namespaces; each question starts a fresh single-turn session that
retrieves only from its namespace, yielding 642 memory-augmented
requests.
$\tau^2$-bench~\cite{barres2025tau} evaluates tool-using customer-service agents in the
airline, retail, and telecom domains. We executed its 178 training tasks with
memory recall disabled and imported the resulting trajectories into OpenViking,
whose extraction pipeline produces customer-scoped records visible only to the
corresponding customer and agent-scoped records recallable by other sessions of
the same domain agent. We then executed the 100 test tasks (20 airline, 40
retail, 40 telecom) as multi-turn sessions with per-turn recall. Each session
contributes one initial warm recall plus its logged per-turn recalls, giving 219
memory-augmented requests (44, 86, 89) drawn from the 533 test-split agent
requests. GPT-5-mini~\cite{gpt5mini} drives the agent, the user simulator, and
memory extraction. Both workloads use frozen
memory snapshots during measurement; memory writes, updates, and invalidation
are outside the scope of this characterization.

\noindent\textbf{Prompt construction and controls.}
Each retrieval returns at most $K{=}15$ records. Following OpenClaw's native
prompt construction, retrieved records appear immediately after the system
prompt in LoCoMo and, in $\tau^2$, after the accumulated interaction history
within the newest user turn or tool result. We render prompts with the Qwen3
chat template and tokenize them with the Qwen3-8B tokenizer~\cite{qwen3technicalreport}, and
enable deferred tool loading so that unused tool definitions are not repeated in
every request~\cite{fei2025mcp,openai_tool_search_docs,anthropic_tool_search_docs}; the $\tau^2$ agent carries three
dispatch tools and reaches domain tools through search. Before tokenization we replace request-specific session identifiers and wall-clock
timestamps with fixed placeholders. This normalization prevents request-unique
metadata alone from terminating an otherwise reusable prefix, thereby giving
exact-prefix caching a favorable baseline; all other prompt content remains
unchanged.

\noindent\textbf{Token classification.}
A \emph{stream} is one LoCoMo conversation or one $\tau^2$ domain, matching the
visibility scope of its records. The \emph{matching prefix} of a request is the longest token-level prefix it shares with any
earlier prompt in the same stream; assuming that all earlier prompts remain
available makes this an optimistic upper bound on exact-prefix reuse. We assign
every token of a request's newly injected memory records to one of three
disjoint categories: \emph{prefix-covered} if it falls within the matching
prefix; a \emph{non-prefix repeat} if it falls beyond the matching prefix and
belongs to a record whose token sequence already appeared in an earlier prompt
of the same stream; and \emph{new} otherwise.

\vspace{-3.5ex}
\subsection{Recalled Memories Recur Outside Prefix}
\label{sec:agent_workload:where}
\vspace{-1.5ex}
\begin{figure}[!t]
    \centering
    \begin{subfigure}[t]{0.48\linewidth}
        \centering
        \includegraphics[width=\linewidth,trim={0 -18.755bp 0 0},clip]{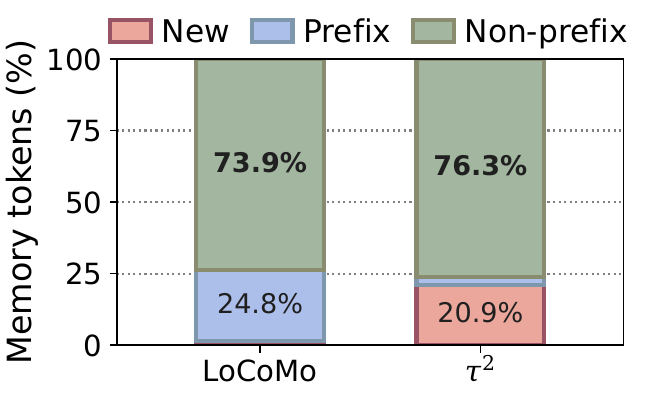}
        \vspace{-4ex}
        \caption{Memory token composition}
        \label{fig:reuse-in-mem}
    \end{subfigure}
    \hfill
    \begin{subfigure}[t]{0.48\linewidth}
        \centering
        \includegraphics[width=\linewidth,trim={0 0 -16.873bp -22.310bp},clip]{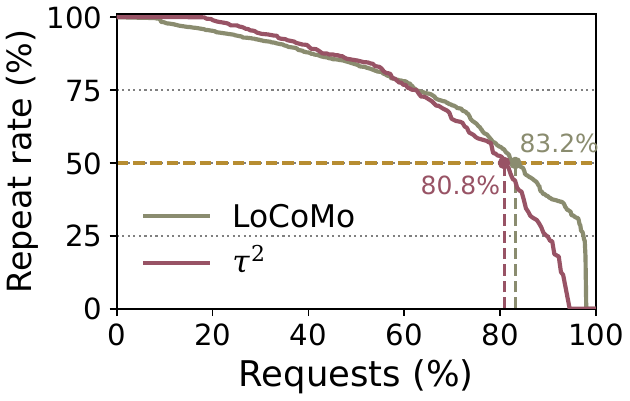}
        \vspace{-4ex}
        \caption{Non-prefix repeat rate}
        \label{fig:reuse-mem-across-reqs}
    \end{subfigure}
    \vspace{-2ex}
    \caption{Characteristics of injected memory tokens and their recurrence across agent requests.}
    \vspace{-5ex}
    \label{fig:agent-memory-mem}
\end{figure}

We first quantify recurrence within newly injected memory records, while determining the matching prefix from the complete prompt. For each workload, Figure~\ref{fig:reuse-in-mem} aggregates token counts across the analyzed requests and reports each category as a fraction of all tokens in these memory records. Of these memory-record tokens, 73.9\% in LoCoMo and 76.3\% in $\tau^2$ repeat beyond the matching prefix. These shares quantify potential exact recurrence missed by prefix-keyed reuse; they are not cache hit rates, realized prefill savings, or guarantees that the corresponding KV states remain resident. Repeats that prefix caching does cover are a minority in LoCoMo (24.8\%) and nearly absent in $\tau^2$ (2.8\%), and new content accounts for the remainder. The two workloads reach this outcome by different routes. In LoCoMo, the memory block directly follows the system prompt, so a change in the leading records or their order breaks the matching prefix, leaving subsequent records outside it. In $\tau^2$, the block follows the conversation history and the new user turn or tool result, whose differences across requests typically cause the prefix to diverge before reaching the recalled records.

Figure~\ref{fig:reuse-mem-across-reqs} shows that this opportunity is typical rather than driven by a small number of requests. Specifically, 83.2\% of LoCoMo requests and 80.8\% of $\tau^2$ requests have at least half of the tokens in their injected memory records classified as repeated beyond the matching prefix.

\begin{figure}[t]
    \centering
    \begin{subfigure}[t]{0.48\linewidth}
        \centering
        \includegraphics[width=\linewidth]{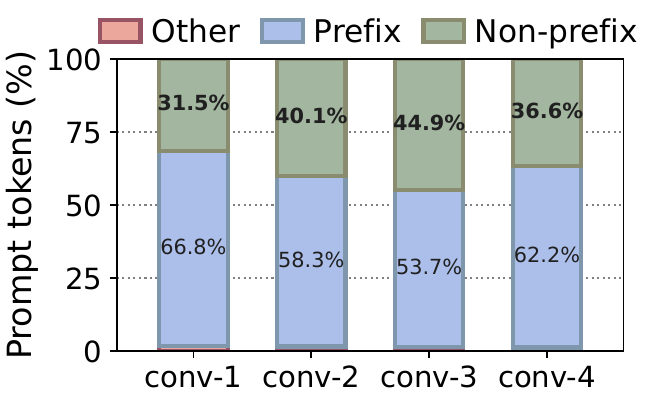}
        \vspace{-3ex}
        \caption{LoCoMo conversations}
        \label{fig:prefill-locomo}
    \end{subfigure}
    \hfill
    \begin{subfigure}[t]{0.48\linewidth}
        \centering
        \includegraphics[width=\linewidth]{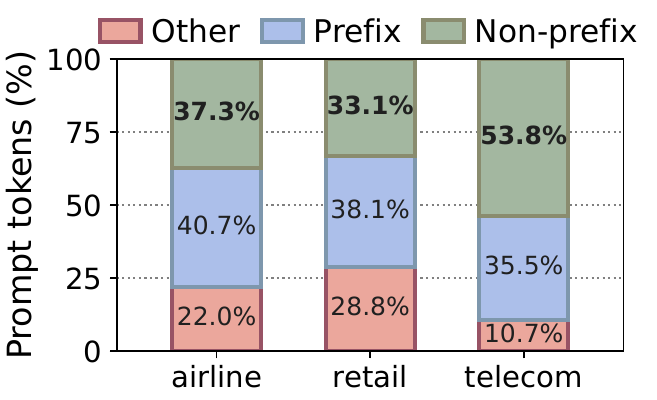}
        \vspace{-3ex}
        \caption{$\tau^2$ domains}
        \label{fig:prefill-tau2}
    \end{subfigure}
\vspace{-3ex}
    \caption{Complete-prompt token composition per stream: matching prefix, repeated memory beyond it, and other tokens.}
    \label{fig:prefill-composition}
    \vspace{-4ex}
\end{figure}

This opportunity remains material when measured against the complete prompt. As shown in Figure~\ref{fig:prefill-composition}, repeated memory beyond the matching prefix accounts for 31.5--44.9\% of complete prompt tokens across the four LoCoMo conversations and 33.1--53.8\% across the three $\tau^2$ domains. In every stream, it also constitutes more than
half of all tokens outside the matching prefix. Thus, the high recurrence within recalled memory identified in Figure~\ref{fig:agent-memory-mem} represents a substantial reuse opportunity at the full-prompt level.

\noindent\textbf{Takeaway.} Non-prefix memory recurrence is both common and substantial, making the context unit, not the prompt prefix, the natural unit of reuse.
\vspace{-3ex}
\section{\sys Overview}
\label{sec:overview}
\vspace{-2.5ex}
\begin{figure*}[t]
    \centering
    \includegraphics[width=0.93\linewidth]{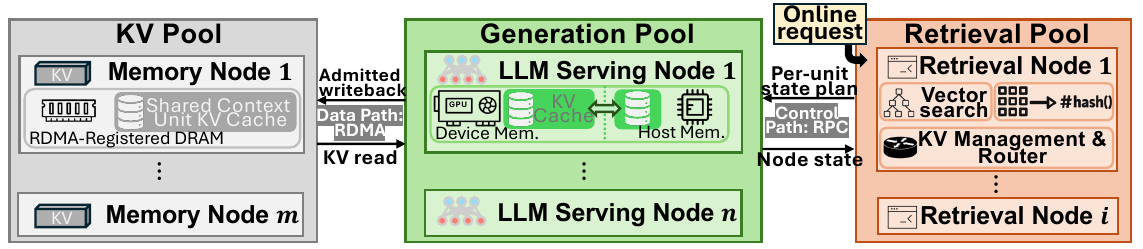}
    \vspace{-2.5ex}
    \caption{\sys design overview.}
    \vspace{-4.5ex}
    \label{fig:design:overview}
\end{figure*}

\sys is a disaggregated serving architecture for online, demand-driven reuse of recurring context units in RAG and retrieval-based agent memory. \sys organizes KV reuse around two decisions for each retrieved context unit: how to prepare its state for the current request, and whether to retain that state for future requests. Retrieval exposes the context units required by a request before model execution, allowing \sys to coordinate these decisions with KV residency and generation load. \sys separates retrieval and reuse planning, shared KV memory, and model execution into independently provisioned Retrieval, KV, and Generation Pools, as shown in Figure~\ref{fig:design:overview}.

\noindent\textbf{Request flow.} An online request enters the Retrieval Pool, which retrieves and identifies its context units. Using retrieval history, KV residency, and generation load, \sys selects a generation node and produces a per-unit preparation and retention plan. The selected node obtains each unit from its local cache, the shared KV Pool, or independent prefill, overlapping these paths. Once the states are ready, it composes them and continues generation. Newly computed states may be retained locally or written back to the KV Pool asynchronously.

\noindent\textbf{Retrieval Pool.}
The Retrieval Pool executes retrieval to obtain the ordered context units required by each request and coordinates their KV reuse. It selects a generation node and produces a per-unit plan specifying how each state should be prepared and whether it should be retained for future reuse. It makes these decisions by combining retrieval demand with KV residency and generation load (Sections~\ref{sec:design:interface} and~\ref{sec:design:policy}).

\noindent\textbf{Generation Pool.}
Each generation node runs an LLM serving engine with a context-unit cache in pinned host memory. It executes the per-unit plan by copying local states, fetching shared states, and prefilling missing units concurrently while continuing to serve other requests. Once the required states are ready, the node recomputes the tokens selected by the composition policy and processes the remaining prompt before decoding (Section~\ref{sec:design:execution}). Authorized states
are retained locally or written back to the shared KV Pool asynchronously, without delaying the current request
(Section~\ref{sec:design:retention}).

\noindent\textbf{KV Pool.}
The KV Pool provides shared, RDMA-accessible memory for context-unit states, extending reuse across generation nodes. The Retrieval Pool manages residency metadata and storage allocation, while generation nodes transfer KV payloads directly to and from memory nodes. Section~\ref{sec:design:retention} describes the consistency and lifetime mechanisms that make these transfers safe.
\vspace{-3ex}
\section{\sys Design}
\label{sec:design}
\vspace{-3ex}
This section describes how \sys turns retrieval-time context-unit demand into cache-management, routing, and execution decisions. We first describe the per-unit planning interface that connects retrieval demand to KV reuse (Section~\ref{sec:design:interface}). We then address the three serving challenges identified in Section~\ref{bg:problem}: coordinating cache management with request routing (Section~\ref{sec:design:policy}), executing requests with partially available states (Section~\ref{sec:design:execution}), and retaining newly computed states off the request's critical path (Section~\ref{sec:design:retention}).

\vspace{-3ex}
\subsection{Retrieval-Coupled KV Interface}
\label{sec:design:interface}
\begin{figure}[t]
    \centering
    \includegraphics[width=0.85\columnwidth]
        {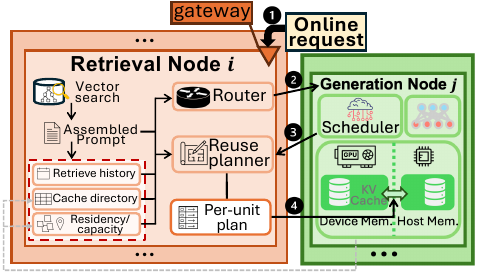}
    \vspace{-2.5ex}
    \caption{The retrieval-side control plane.}
    \vspace{-5ex}
    \label{fig:design:retrieve_panel}
\end{figure}

\vspace{-2ex}
Once retrieval determines the context units for a request, \sys must turn this result into an execution plan for a specific generation node. The plan must identify each unit consistently across requests, specify where its state should
come from, and carry any permission to retain that state after preparation. \sys exposes these decisions through a per-unit interface shared by the Retrieval and Generation Pools. Figure~\ref{fig:design:retrieve_panel} illustrates this control flow.

\noindent\textbf{Stable context-unit identity.}
Within each sharing scope, \sys assigns a context unit an identity derived from its token sequence and serving configuration. The identity is independent of
the unit's position or surrounding context, allowing retrieval demand and cached KV states to refer to the same unit across requests. The Retrieval Pool uses this identity to track retrieval history and KV residency.

\noindent\textbf{Node-specific planning.}
Preparation depends on where the request executes: a state that is local on one generation node may require a remote fetch or computation on another. \sys therefore selects a generation node and dispatches the request to it (Step~\fcirc{2} of Figure~\ref{fig:design:retrieve_panel}) before finalizing the
per-unit plan. The Retrieval Pool's directory tracks residency across nodes and drives this selection; the node returns its current cache snapshot (Step~\fcirc{3}), which revalidates local availability before the plan is finalized.
Section~\ref{sec:design:policy} describes node selection and cache management.

\noindent\textbf{Per-unit plan.}
Using the selected node's snapshot and shared-cache residency, the Retrieval Pool produces a plan for each retrieved unit and returns it to the node (Step~\fcirc{4}). The preparation field specifies whether the node should use a local state, fetch a state from the shared KV Pool, or compute the state from text. Retention is specified separately: the plan may authorize the resulting state to be kept in the node's local cache, written back to the shared KV Pool, both, or neither. Thus, a unit's availability for the current request does not depend on whether \sys chooses to retain it for future reuse. What is retained is the unit's reusable form, its keys before RoPE together with its values, which is what both cache tiers store and transfer.

\noindent\textbf{Storage blocks.}
Context units vary in length, so both cache tiers store their states in fixed-size blocks of up to $B$ tokens. A unit of $n$ tokens occupies $\lceil n/B \rceil$ blocks, which need not be adjacent in memory and, in the shared KV Pool, may reside on different memory nodes. \sys manages reuse at unit granularity. Admission and eviction always operate on whole context units, while blocks provide the granularity for capacity allocation and data transfer. A unit becomes readable only after all of its blocks have been written.

\vspace{-4ex}
\subsection{Retrieval-Aware Caching and Routing}
\label{sec:design:policy}
\vspace{-2ex}

Cache placement and request routing determine each other's value: retaining a context unit on a generation node is useful only if future requests for that unit execute there, while routing requests solely for cache locality can overload the node holding popular states. \sys uses retrieval history to decide which KV states to retain, and routes requests based on local KV residency and generation load.

\noindent\textbf{Estimating reuse value.}
Retrieval demand changes over time, but some context units remain popular across longer periods. For a context unit $u$ at request ordinal $t$, \sys scores reuse using both recent and cumulative retrieval frequency:
\[
\setlength{\abovedisplayskip}{2pt}
\setlength{\belowdisplayskip}{2pt}
\setlength{\abovedisplayshortskip}{2pt}
\setlength{\belowdisplayshortskip}{2pt}
    s(u,t)
    = w_r f_W(u,t) + w_\ell f_\infty(u,t),
\]
where $f_W(u,t)$ is the number of requests that retrieved $u$ among the last $W$ requests, and $f_\infty(u,t)$ is its cumulative retrieval count. We fix $w_r$ = 1 as the unit of the score. $W$ is set to the number of recent requests whose retrieved units would fill the cache, so that $f_W$ measures demand over the horizon the cache can retain; it thus follows from cache capacity and the average number of units per request rather than being tuned per workload. $w_\ell$ = 0.05 is held fixed across all experiments.

The two cache tiers apply the same scoring rule to different request streams. The shared cache uses retrievals from all generation nodes, whereas each local cache uses only requests routed to that node. Thus, the same context unit may receive different scores in the local and shared caches.

\noindent\textbf{Admission and eviction.}
A unit becomes an admission candidate once its score reaches a threshold $\theta$. By default, a unit is not admitted after its first retrieval, preventing one-time accesses from consuming retained capacity. If capacity is insufficient, \sys considers resident units in increasing score order and admits the candidate only if units with strictly lower scores can free enough space. Replacement operates on whole context units, although capacity is accounted for in blocks. Units whose demand fades lose their
recent-history contribution and can therefore be displaced by more frequently retrieved units. Section~\ref{sec:design:retention} describes when a selected
victim is safe to reclaim.

\noindent\textbf{Cache-aware routing.}
Cache locality should not come at the cost of overloading a generation node. \sys first restricts consideration to nodes whose in-flight request count is within one request of the least-loaded node. Among these nodes, it selects the
one holding the largest number of local blocks from the request's context units, breaking ties by lower load and then round-robin. Shared-cache hits do not affect this ranking because they are accessible from every generation node. The load constraint prevents popular locally cached units from drawing too many requests to the same node.

Local and shared admission are evaluated independently using their respective demand histories. A missing unit may therefore be retained locally, in the shared KV Pool, in both, or in neither, while a remote hit may still be admitted locally. The resulting retention permissions are carried in the per-unit plan described in Section~\ref{sec:design:interface}.

\vspace{-4ex}
\subsection{Executing Partially Cached Requests}
\label{sec:design:execution}
\vspace{-2ex}

\begin{figure}[t]
    \centering
    \includegraphics[width=0.99\columnwidth]
        {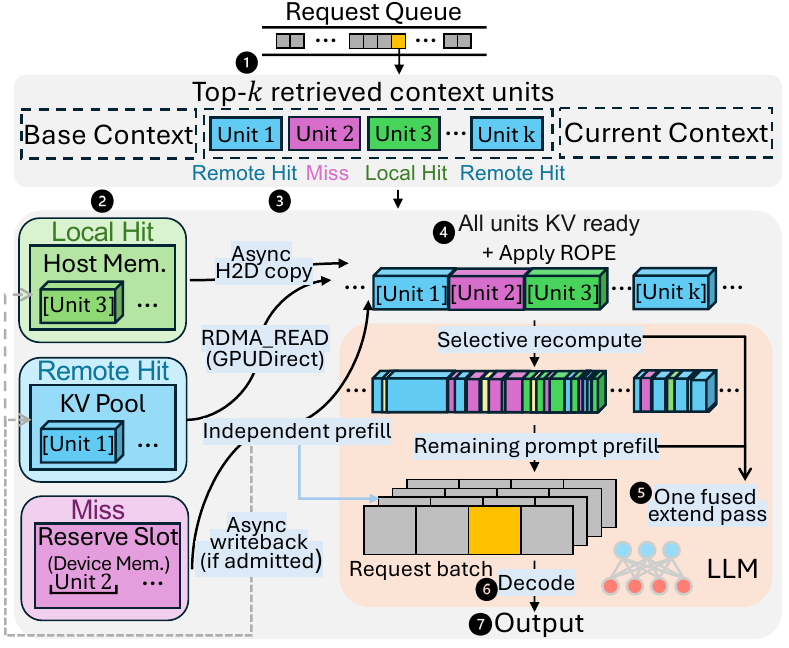}
    \vspace{-2.5ex}
    \caption{Execution of a partially cached request.}
    \vspace{-6ex}
    \label{fig:design:execution}
\end{figure}

A request may contain local hits, remote hits, and misses, whose states become ready at different times. \sys prepares these states independently, transferring cache hits asynchronously while misses enter the serving engine as prefill work. The request becomes ready once all required states are available, without blocking the scheduler from serving other requests. Figure~\ref{fig:design:execution} shows this execution path.

\noindent\textbf{Placement before preparation.}
Before preparing any unit, \sys fixes the prompt layout and allocates its final GPU KV slots. Local copies, remote fetches, and miss prefills can proceed independently without waiting for one another. Separators and other request-specific tokens are processed with the remaining prompt after the unit states are ready.

\noindent\textbf{Preparing cache hits.}
Local and remote hits follow the transfer paths in Step~\fcirc{2} of Figure~\ref{fig:design:execution}. A local hit is copied from the node's pinned host-memory cache, while a remote hit is fetched from the shared KV Pool with one-sided RDMA. Remote blocks pass through bounded GPU staging space before being placed into the request's assigned KV slots. After applying RoPE to the keys at their new positions, \sys marks the unit ready and releases any remote read lease.

\noindent\textbf{Preparing misses through the scheduler.}
Missing units are independently prefilled (Step~\fcirc{3}) from position zero, without a preceding prefix. \sys submits this work as an internal request so that miss prefill joins regular extend batches rather than blocking the parent request.
This independent prefill produces the unit's reusable form. \sys writes it into the unit's assigned slots with RoPE applied at the request's own positions, and, when the plan authorizes retention, keeps the same unrotated copy for the cache, avoiding a second prefill solely to populate it. Concurrent requests on the same generation node that miss on the same unit share one in-flight prefill. This coalescing is local to a generation node; different
nodes may still compute the same missing unit independently.

\noindent\textbf{Completing prefill.}
Once all unit states are ready (Step~\fcirc{4}), the parent request re-enters the scheduler. \sys recomputes token ranges selected by the composition policy and processes the remaining prompt, including separators and the query, in the final extend pass (Step~\fcirc{5}). The recomputed states are used only by the current request. Cached context-unit states remain in their reusable form.

\noindent\textbf{Preparation failures.}
If a cache transfer fails or times out, \sys releases the affected resources and falls back to computing the unit from text. Under tensor parallelism, all ranks agree on the fallback before execution continues, preventing ranks from
taking different preparation paths.

\subsection{Nonblocking State Retention}
\label{sec:design:retention}
\vspace{-2ex}

Retaining a newly computed state benefits future requests, not the current one. \sys therefore performs retention off the request's critical path, while ensuring that a state is never read before it is complete and never reclaimed while a transfer still depends on it.

\noindent\textbf{Reserving a shared entry.}
When a missing unit passes shared admission, the Retrieval Pool reserves its blocks and creates a \textsc{pending} entry tagged with a new storage epoch, a version number that distinguishes successive allocations for the same unit. The generation node receives permission to write the reserved blocks under this storage epoch, and only that writer may publish the entry. A \textsc{pending} entry is not a remote hit. Other requests that need the unit compute it for their own execution without shared-write permission, so no request depends on the first writer finishing.

\noindent\textbf{Asynchronous retention.}
Once a unit's reusable form is available on the generation node, whether fetched or
computed, authorized local insertion and shared writeback proceed independently
of the parent request. For shared writeback, the generation node writes the
reserved blocks with one-sided RDMA. The prototype bounds pinned-host
staging and the number of outstanding writes so that background writeback does
not consume unbounded transfer resources. Source buffers remain valid until
their dependent copies or transfers complete.

\noindent\textbf{Publishing complete states.}
A shared state becomes readable only after all of its reserved blocks have been
written successfully. The writer requests publication after observing RDMA
completion for every block, and the Retrieval Pool verifies the storage
epoch, write permission, and expected block set before atomically changing the
entry from \textsc{pending} to \textsc{ready}. These checks prevent stale
writers or partial writeback from publishing a cache entry.

\noindent\textbf{Protecting readers and reclamation.}
A remote hit carries a read lease on a specific storage epoch. The
referenced blocks cannot be reclaimed while the lease is active, including by
admission decisions made later in the same per-unit plan, and the generation node releases the lease after the transfer completes. Local admission plans are similarly revalidated against current cache state before execution. Thus, routing and admission may use residency as a policy signal, but execution relies only on leases and revalidated entries.

\noindent\textbf{Retention failures.}
A failed writeback is never published. The generation node cancels the
reservation, and its blocks are reclaimed only after outstanding transfers
have completed. Because the current request has already obtained the state it
needs, retention failure only loses a future reuse opportunity; failures that
prevent state preparation instead trigger the fallback in
Section~\ref{sec:design:execution}. Control plane failover and metadata recovery
are discussed in Section~\ref{discussion}.
\vspace{-4ex}
\section{Evaluation}
\label{sec:evaluation}
\vspace{-3ex}

We evaluate \sys by asking three questions:
(1)~Does \sys reduce TTFT across RAG and agent-memory workloads as request rate increases (Sections~\ref{sec:eval:rag} and~\ref{sec:eval:agent})?
(2)~How much do retrieval-aware cache management and cache-aware routing contribute to these gains (Section~\ref{sec:eval:ablation})?
(3)~Does \sys preserve the answer quality of the underlying composable KV-reuse policy (Section~\ref{sec:eval:quality})?

\subsection{Methodology}
\label{sec:eval:methodology}

\noindent\textbf{Implementation.}
We implement \sys on SGLang, extending its scheduler, model forward path, and attention metadata to support context-unit preparation and composition. Missing units are represented as internal requests and scheduled in the same extend batches as regular model work; concurrent misses for the same unit on a generation node share one in-flight preparation. During these prefills, the forward path captures pre-RoPE keys and values for subsequent reuse. For
composition, EPIC's LegoLink policy identifies recomputation ranges, encoded in the attention metadata and executed together with the remaining prompt in a single extend pass. A fused Triton kernel applies RoPE when cached
keys are placed at their request-specific positions. The KV Pool is implemented as a standalone service backed by RDMA-registered host memory. Each memory node exposes fixed-size blocks for one-sided access, while the Retrieval Pool maintains allocation and residency metadata. Generation nodes perform one-sided RDMA transfers, using GPUDirect RDMA when available and pinned host memory otherwise; RDMA completions and CUDA events synchronize state availability and buffer reuse. This testbed exercises cross-node reuse between the generation server and a separate KV server, but places all four generation replicas on one machine. Routing effects at larger replica counts, and generation spread across multiple servers, remain to be validated at scale.

\noindent\textbf{Testbed.}
We develop and evaluate the prototype of \sys on CloudLab~\cite{cloudlab}. Our testbed consists of two Dell PowerEdge XE8545 servers, each with two 24-core AMD EPYC 7413 CPUs, 512 GB host memory, four 40\,GB NVIDIA A100 GPUs, and a Mellanox ConnectX-6 100Gb NIC. One server hosts the Generation
Pool; the other hosts the Retrieval Pool and the KV Pool. We assign each generation node a disjoint host-memory region to emulate isolation across nodes.

\noindent\textbf{Models.} For RAG, we evaluate Llama-3.1-8B-Instruct with four TP=1 replicas and Qwen3-14B with two TP=2 replicas, using four GPUs in both configurations. For agent memory, we use Qwen3-32B with TP=4 and thinking disabled. 

\noindent\textbf{Workloads.} We evaluate \sys on RAG and retrieval-based agent-memory workloads.
\begin{list}{\textbullet}{%
    \setlength{\leftmargin}{1.2em}%
    \setlength{\labelwidth}{0.8em}%
    \setlength{\labelsep}{0.4em}%
    \setlength{\topsep}{0pt}%
    \setlength{\itemsep}{0pt}%
    \setlength{\parsep}{0pt}%
    \setlength{\partopsep}{0pt}%
}
    \item \textbf{RAG.} We evaluate four QA datasets: MMLU~\cite{hendryckstest2021}, a multiple-choice knowledge benchmark; Natural Questions (NQ)~\cite{kwiatkowski-etal-2019-natural}, which contains real Google Search queries; 2WikiMQA~\cite{ho-etal-2020-constructing}, a multi-hop QA benchmark over Wikipedia; and TriviaQA~\cite{joshi-etal-2017-triviaqa}, a knowledge-intensive QA benchmark. Following RAGCache~\cite{jin2025ragcache}, we construct a popularity-selected corpus of 300,000 Wikipedia articles and retrieve the top five documents using FAISS~\cite{faiss}, a widely used vector search engine, with HNSW~\cite{hnsw} indexing and precomputed \texttt{text-embedding-3-small} query embeddings~\cite{openai_embedding_models}. For each dataset, we split the request sequence 1:1 into a warm-up half and a measured half; only the latter is used for reported results. We measure time to first token (TTFT) end to end. TTFT includes online retrieval and generation. Query embeddings are precomputed and identical across systems, so we exclude embedding computation and measure from retrieval onward.

    \item \textbf{Agent memory.} We replay LoCoMo requests across four conversations and $\tau^2$ requests across three domains, following Section~\ref{sec:agent_workload:method}. The trajectories were generated with GPT-5-mini~\cite{gpt5mini} driving the agent, simulated user, and memory extraction, and are replayed unchanged on our testbed. For each LoCoMo conversation and each $\tau^2$ domain, we use the first 10\% of requests for warm-up and report results over the remaining 90\%. Each request therefore follows a fixed recorded trajectory, so later turns do not depend on the serving model's outputs. Requests are issued through OpenClaw with OpenViking-backed memory, using a separate OpenClaw instance for each conversation or domain and a shared generation backend. Both workloads use frozen memory snapshots.

\end{list}

\begin{figure*}[t]
    \centering
    \includegraphics[width=0.9\textwidth]{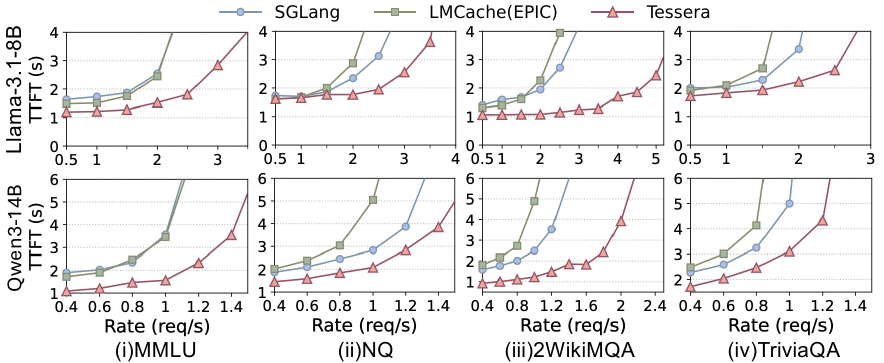}
    \vspace{-2.5ex}
    \caption{Mean TTFT vs.\ RAG request rate.}
    \label{fig:rag-main-result}
    \vspace{-4ex}
\end{figure*}

\noindent\textbf{Baselines.}
We compare \sys with two baselines.

\begin{list}{\textbullet}{%
    \setlength{\leftmargin}{1.2em}%
    \setlength{\labelwidth}{0.8em}%
    \setlength{\labelsep}{0.4em}%
    \setlength{\topsep}{0pt}%
    \setlength{\itemsep}{0pt}%
    \setlength{\parsep}{0pt}%
    \setlength{\partopsep}{0pt}%
}
    \item \textbf{SGLang.}
    We use unmodified SGLang~\cite{zheng2024sglang} at commit \texttt{f81dec0}, the same commit on which \sys is built. Prefix caching reuses KV states only for matching prompt prefixes; context units beyond the matching prefix are prefilled from text.

    \item \textbf{LMCache (EPIC).}
    LMCache~\cite{liu2025lmcache} provides a general KV storage and transfer layer, while its CacheBlend integration assumes that reusable context-unit KV states are precomputed and available before serving, rather than managing their residency dynamically as they recur online. To isolate the serving architecture from the composition policy, we replace CacheBlend's selective-recomputation policy with EPIC's LegoLink~\cite{hu2025epic}, which \sys also uses, and use the same recomputation configuration in both systems. We extend the integration with the KV-transfer support required by our configuration but otherwise leave LMCache's storage and transfer architecture unchanged. The baseline runs on vLLM~\cite{kwon2023efficient}, using LMCache's native local host-memory cache and Mooncake~\cite{qin2024mooncake} backend for shared remote KV storage.
\end{list}

\noindent\textbf{Experimental protocol.}
Requests follow a Poisson arrival process at the reported aggregate rate. Unless otherwise noted, each request generates one output token, allowing us to isolate the prefill-side effects of KV reuse on TTFT and queueing. This setting also isolates the work that would remain on the prefill side of a prefill--decode disaggregated deployment. Given the size of our testbed, we do not separately provision a decode pool and therefore do not evaluate
long-decode throughput. For RAG, TTFT is measured from the start of retrieval until the first output token and excludes query-embedding computation. For agent memory, TTFT is measured from the client's HTTP request to OpenClaw
until the first nonempty content arrives in the response stream.

\subsection{RAG Serving Performance}
\label{sec:eval:rag}
\vspace{-2ex}

We evaluate \sys against SGLang and LMCache (EPIC) in terms of mean TTFT across four RAG datasets, increasing request rates, and two model configurations. 

Figure~\ref{fig:rag-main-result} shows that \sys consistently lowers mean TTFT and sustains low latency to higher request rates across all four RAG workloads and both model configurations. The difference is modest at low request rate, where queueing is limited, and grows as prefill becomes a bottleneck. On 2WikiMQA with Llama-3.1-8B-Instruct, \sys achieves approximately 1.1\,s mean TTFT at 2.5 req/s, compared with 2.7\,s for SGLang and approximately 4\,s for LMCache (EPIC), a reduction of up to 3.6$\times$. On MMLU with Qwen3-14B, \sys achieves approximately 1.6\,s at 1 req/s, versus approximately 3.5\,s for the baselines. The gap is smaller at low request rate, particularly on NQ, and all systems eventually exhibit rapidly increasing TTFT as request rate rises. The widening gap at higher request rates reflects reduced prefill work: requests spend less GPU time reconstructing recurring context states, delaying the onset of queueing under the evaluated workload. LMCache (EPIC) has higher mean TTFT than SGLang at most request rates because its integration assumes a precomputed KV store rather than online state construction. Thus, our comparison evaluates the extended integration (Section~\ref{sec:eval:methodology}).

\vspace{-3ex}
\subsection{Agent-Memory Serving Performance}
\label{sec:eval:agent}
\vspace{-2ex}

\begin{figure*}[t]
    \centering
    \begin{minipage}[t]{0.48\textwidth}
        \centering
        \includegraphics[width=\linewidth]{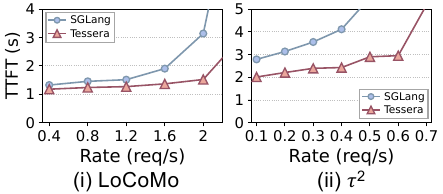}
        \vspace{-5ex}
        \caption{Mean TTFT vs.\ agent-memory request rate.}
        \vspace{-3ex}
        \label{fig:agent-load}
    \end{minipage}
    \hfill
    \begin{minipage}[t]{0.48\textwidth}
        \centering
        \includegraphics[width=\linewidth]{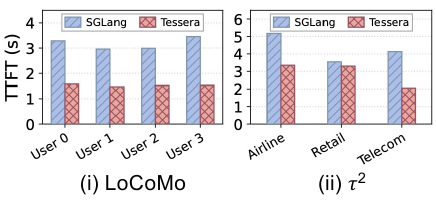}
        \vspace{-5ex}
        \caption{Mean TTFT across users and task domains.}
        \vspace{-3ex}
        \label{fig:agent-ttft}
    \end{minipage}
\end{figure*}


\begin{figure*}[t]
    \centering
    \begin{minipage}[b]{0.32\textwidth}
        \centering
        \includegraphics[width=\linewidth]{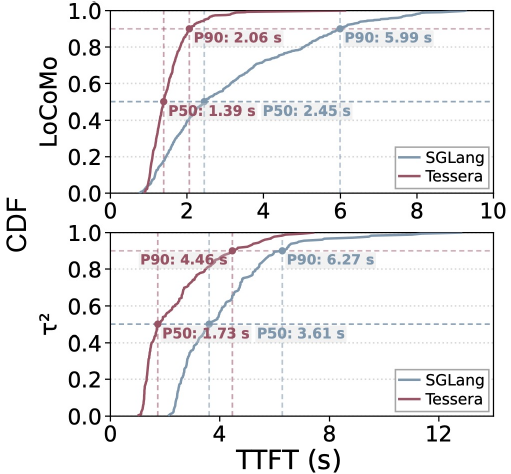}
        \vspace{-5ex}
        \caption{CDF of per-request TTFT for the agent workloads.}
        \vspace{-4ex}
        \label{fig:agent-ttft-cdf}
    \end{minipage}\hfill
    \begin{minipage}[b]{0.32\textwidth}
        \centering
        \includegraphics[width=\linewidth]{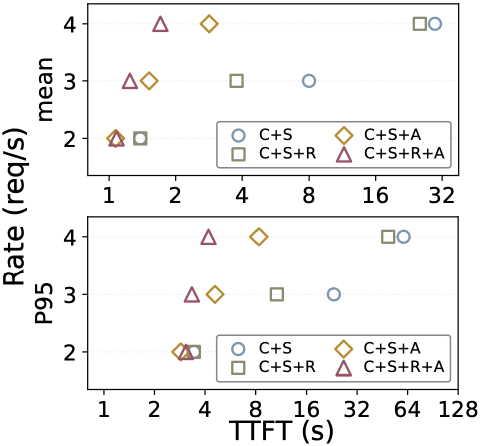}
        \vspace{-23pt}
        \caption{Ablation on 2WikiMQA: (i) mean TTFT and (ii) P95 TTFT.}
        \vspace{-4ex}
        \label{fig:ablation-result}
    \end{minipage}\hfill
    \begin{minipage}[b]{0.32\textwidth}
        \centering
        \includegraphics[width=\linewidth]{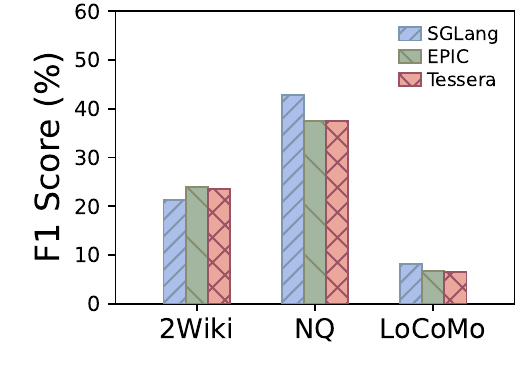}
        \vspace{5pt}
        \caption{Answer quality (F1) under selective recomputation.}
        \vspace{-4ex}
        \label{fig:quality}
    \end{minipage}
    \vspace{-1ex}
\end{figure*}

We evaluate \sys against SGLang on LoCoMo and $\tau^2$. Figure~\ref{fig:agent-load} shows lower mean TTFT with \sys on both workloads, with the gap widening as offered load increases. At 2.0 req/s on LoCoMo, \sys achieves approximately 1.5\,s, compared with 3.1\,s for SGLang. At 0.4 req/s on $\tau^2$, it achieves approximately 2.4\,s, compared with 4.1\,s. With \sys, the sharp rise occurs at a higher offered load, but it is not eliminated.

Figures~\ref{fig:agent-ttft} and~\ref{fig:agent-ttft-cdf} examine the same representative request rate by user and across requests. Every LoCoMo conversation sees a 2.0--2.3$\times$ speedup in mean TTFT. The $\tau^2$ speedup ranges from 1.07$\times$ for retail to 2.0$\times$ for telecom. On LoCoMo, median and P90 TTFT fall from 2.45\,s and 5.99\,s to 1.39\,s and 2.06\,s, respectively. On $\tau^2$, median TTFT falls from 3.61\,s to 1.73\,s, while P90 improves more modestly from 6.27\,s to 4.46\,s. These measurements cover individual model requests issued by the agent, not end-to-end agent-task completion.

\vspace{-4ex}
\subsection{Effect of Cache Management and Routing}
\label{sec:eval:ablation}
\vspace{-2ex}

We isolate cache admission/eviction and request routing on 2WikiMQA with four Llama-3.1-8B-Instruct TP=1 replicas. All configurations use composable reuse (C), shared storage (S), and the same execution path. We independently ablate the retrieval-aware admission/eviction (A) and cache-aware routing (R) mechanisms described in Section~\ref{sec:design:policy}. Figure~\ref{fig:ablation-result} reports mean and P95 TTFT for four configurations:
 \vspace{-1ex}
\begin{list}{\textbullet}{%
\setlength{\leftmargin}{1.2em}%
\setlength{\labelwidth}{0.8em}%
\setlength{\labelsep}{0.4em}%
\setlength{\topsep}{0pt}%
\setlength{\itemsep}{0pt}%
\setlength{\parsep}{0pt}%
\setlength{\partopsep}{0pt}%
}
\item \textbf{C+S (Baseline):} LRU eviction and load-aware routing without cache awareness;
\item \textbf{C+S+R (+Routing):} load- and cache-aware routing with LRU eviction;
\item \textbf{C+S+A (+Policy):} retrieval-aware admission and eviction with load-aware routing;
\item \textbf{C+S+R+A (Full):} both policies, representing the full \sys.
\end{list}
 \vspace{-1ex}
Admission and eviction have the larger effect at higher request rates. At 4 req/s, enabling A reduces mean TTFT from 29.7\,s to 2.8\,s, whereas routing alone leaves latency above 25\,s. Adding cache-aware routing to A further reduces mean and P95 TTFT by approximately 40\% and 50\%, respectively. At the lowest request rate, the configurations are much closer. At the two higher request rates, combining routing with retrieval-aware admission further reduces both mean and P95 TTFT, showing that locality-aware routing is useful once the cache retains the units that recur under load. Because composable reuse, shared storage, and the execution path are fixed across configurations, this experiment isolates the two online management policies introduced in Section~\ref{sec:design:policy}.

\vspace{-3ex}
\subsection{Answer Quality}
\label{sec:eval:quality}
\vspace{-2ex}

\sys does not introduce a new composition policy; our prototype uses EPIC's policy. We therefore ask whether \sys's serving path introduces additional quality loss relative to EPIC. We evaluate whether \sys preserves the answer quality of EPIC~\cite{hu2025epic}. We compare \sys with EPIC's reference implementation and include SGLang full prefill as a reference. We report answer F1 against gold answers on a 0--100 scale. The evaluation uses 500 questions each from 2WikiMQA and NQ and 200 questions from LoCoMo, with the same recomputation budgets for \sys and EPIC.

Figure~\ref{fig:quality} shows only minor quality differences between \sys and EPIC across all three workloads. On 2WikiMQA, NQ, and LoCoMo, \sys achieves F1 scores of 23.54, 37.52, and 6.58, compared with 23.88, 37.57, and 6.70 for EPIC, respectively. The largest gap is only 0.34 F1 points on a 0--100 scale. Both systems also show the same direction of quality change relative to full prefill: higher F1 on 2WikiMQA and lower F1 on NQ and LoCoMo. On LoCoMo, their generated answers achieve 92.42 pairwise F1. Because \sys and the EPIC baseline run on different serving engines, exact generated-text agreement is not expected even under full recomputation. For comparison, full recomputation with identical input tokens across SGLang and vLLM yields 90.83 pairwise F1, showing that cross-framework execution alone can produce non-identical outputs.

\vspace{-3.5ex}
\section{Related Work}
\label{related_work}
\vspace{-2.5ex}

\noindent\textbf{The Evolution of Composable KV Reuse.} 
Prefix-based KV caching works well when overlap appears as a stable contiguous prefix, such as a system prompt or repeated conversation history, and is supported by systems such as Gemini context cache~\cite{google_gemini_context_caching}, vLLM~\cite{kwon2023efficient}, and SGLang~\cite{zheng2024sglang}. However, prefix reuse cannot capture all recurring content in RAG workloads: retrieved document chunks often reappear beyond the matching prefix as retrieval results change in composition and order~\cite{jin2025ragcache}. PromptCache~\cite{gim2024prompt} enables modular reuse through schema-defined position IDs and position-preserving layouts. TurboRAG~\cite{lu2025turborag} concatenates independently precomputed chunks, but removes cross-chunk attention and thus requires fine-tuning to adapt the model. CacheBlend~\cite{yao2025cacheblend} dynamically selects context-sensitive tokens for recomputation, whereas EPIC~\cite{hu2025epic} uses a fixed chunk-initial recomputation window to mitigate attention-sink effects. KVLink~\cite{yang2026kvlink} uses trainable link tokens to connect encoded chunks, enabling cross-chunk context with limited online KV computation. These methods determine how an available unit state is composed into a prompt. \sys addresses the complementary online problem: which unit states exist and where they reside, where a request executes, how cached and missing states are prepared together, and which newly computed states are retained. Our prototype uses EPIC as its composition policy.

\noindent\textbf{RAG-Aware Chunk-Cache Management.}
RAGCache and Cache-Craft also couple retrieval with cache management, but retain states whose validity depends on the prompt prefix or surrounding context. RAGCache~\cite{jin2025ragcache} organizes prefix-dependent document states in a knowledge tree and combines retrieval-aware replacement with cache-aware scheduling. 
Cache-Craft~\cite{agarwal2025cache} removes RoPE from previously computed chunk states but retains their preceding-context dependencies, maintaining context-specific variants in a single-node GPU/CPU/SSD hierarchy while overlapping cache loading, miss prefill, and selective recomputation. HYPIC~\cite{liu2026hypicacceleratinghybridattentionllm} combines segment-level cache lookup with cross-instance miss prefill and state assembly for hybrid-attention models. These systems decide retention for prefix-dependent or context-specific states within a single node's memory hierarchy. \sys instead manages independently prepared context-unit states whose identity remains stable across prompt positions and surrounding contexts, coordinating their demand and residency across generation nodes and shared memory.

\noindent\textbf{Distributed KV-Cache Management and Scheduling.} Prefill--decode disaggregation systems such as DistServe~\cite{zhong2024distserve} and Splitwise~\cite{patel2024splitwise} separate the two inference phases to match their distinct resource demands. Beyond execution disaggregation, Mooncake~\cite{qin2024mooncake}, MemServe~\cite{memserve}, LMCache~\cite{liu2025lmcache}, and InfiniStore~\cite{infinistore} extend KV-cache storage and access across devices and serving nodes through tiered memory, remote transfer, and shared cache pools. Cache-aware schedulers such as Preble~\cite{ICLR2025_5bc342f4}, as well as production KV-aware routers in SGLang, Mooncake, and NVIDIA Dynamo~\cite{nvidia_dynamo_router_operations}, further balance prompt-prefix locality against worker load. \sys builds on these capabilities by making retrieval-selected context units the objects of cluster-wide coordination, linking their demand and residency to cache retention, request routing, and per-request state preparation.

\noindent\textbf{KV Cache Reuse for Agent Workloads.} Prior work exploits recurring structure in agent workloads through lookahead preparation of transformed contexts~\cite{pan2026smoothagentefficientlonghorizonllmbased}, reuse of relatively ordered segments under absolute-position shifts~\cite{315967}, and prefix sharing across same-architecture fine-tuned model variants~\cite{316100}. KVCOMM~\cite{ye2025kvcomm} aligns shared-content caches under diverse agent prefixes, while TokenDance~\cite{tokendance} collectively reuses and compresses shared blocks in synchronized multi-agent rounds. Recent work also examines reuse fidelity: judge-side reuse of candidate-cache blocks can weaken cross-candidate interactions~\cite{liang-etal-2026-kv}, whereas AgentKVShift~\cite{pandey2026agentkvshiftefficientkvcache} applies probe-guided correction to retrieved-memory states. These works target workflow-specific cache preparation, reconstruction, or sharing. \sys instead treats retrieved memory records through the same context-unit interface used for RAG and manages their reuse across requests.
\vspace{-4ex}
\section{Discussion}
\label{discussion}
\vspace{-3ex}

\noindent\textbf{Algorithm and Model Compatibility.} \sys currently supports RoPE-based full-attention models with conventional per-token K/V caches and EPIC recomputation~\cite{hu2025epic}, accommodates other composition policies based on independent context-unit states and token-range recomputation, and requires extensions for hybrid-attention, MLA, or non-RoPE models, which we do not evaluate.

\noindent\textbf{Content Updates and Isolation.}
Source documents may be revised, and agent memories may be updated or consolidated~\cite{zhang2025memorysurvey}. Content-based identities distinguish revised units while preserving reuse of unchanged ones. Multi-tenant deployment requires namespaces aligned with each record's authorized visibility scope, distinguishing customer-private and agent-shared records within tenant boundaries, with authorization checked before KV resolution. Superseded versions may undergo eviction, whereas explicit deletion or access revocation requires invalidating affected shared and local entries, rejecting readmission and stale writeback, and reclaiming storage after outstanding accesses are safely fenced or drained. Our agent characterization uses fixed memory snapshots. Online mutation and revocation remain outside its scope.

\noindent\textbf{Comparison with RAG-specific Cache Systems.}
RAGCache and Cache-Craft are relevant end-to-end RAG serving systems, but neither operates on independently prepared context-unit states. Our controlled comparison therefore uses LMCache with the same EPIC composition policy as \sys, allowing us to compare online cache management while holding the underlying reuse mechanism fixed. Comparing against RAGCache or Cache-Craft end to end would additionally vary how reusable states are constructed and composed.

\noindent\textbf{Prefill--Decode Disaggregation.} \sys's reuse mechanisms operate on prompt preparation and can precede a separate decode stage in a prefill--decode disaggregated deployment~\cite{zhong2024distserve,patel2024splitwise}. Our prototype does not implement this handoff, and we therefore do not characterize long-decode throughput or interactions with decode-side scheduling.

\noindent\textbf{Control-Plane Deployment.}
\sys hosts its control plane on the retrieval nodes rather than as a separate tier, and handles two control RPCs per routed request for route preparation and cache classification. Token caching reduces repeated tokenization, while directory maintenance and routing costs depend on request volume, replica count, and cache occupancy. Our performance claims are limited to the evaluated cluster configurations. Routing quality at larger replica counts, controller throughput under higher request volume, and controller failover with metadata recovery remain open deployment questions. Preble's parallel request processing~\cite{ICLR2025_5bc342f4} and Mooncake's reconstruction of scheduling metadata from authoritative node state~\cite{qin2024mooncake} provide relevant directions for these extensions.

\vspace{-3.5ex}
\section{Conclusion}
\vspace{-3ex}

RAG and retrieval-based agent memory create substantial opportunities for KV reuse beyond the matching prompt prefix; in our agent workloads, 73.9--76.3\% of injected memory tokens are such repeats. \sys couples retrieval demand with KV residency and generation load, while supporting mixed cached/missing execution and nonblocking state retention. \sys reduces mean TTFT by up to 3.6$\times$ and sustains low latency where the baselines saturate, while matching the answer quality of the underlying composition policy.

\vspace{-1ex}

\bibliographystyle{plain}
\bibliography{sample-base}

\end{document}